**Title: Beaming light from a quantum emitter with a planar optical antenna**

**Running Title: Beaming an emitter with a planar optical antenna**

**Authors:**

Simona Checcucci,[1,2,3,4] (e-email: checcucci@lens.unifi.it)

Pietro Lombardi,[1,2,3] (e-mail: lombardi@lens.unifi.it)

Sahrish Rizvi,[1] (e-mail: rizvi@lens.unifi.it)

Fabrizio Sgrignuoli,[1,3] (e-mail: sgrignuoli@lens.unifi.it)

Nico Gruhler,[5,6] (e-mail: nico.gruhler@kit.edu)

Frederik B. C. Dieleman,[7] (e-mail: frederik.dieleman12@imperial.ac.uk)

Francesco Saverio Cataliotti,[1,3,4] (e-mail: fsc@lens.unifi.it)

Wolfram H. P. Pernice,[6] (e-mail: wolfram.pernice@uni-muenster.de)

Mario Agio,[1,2,4,8] (e-mail: mario.agio@uni-siegen.de)

Costanza Toninelli,[1,2,4] (e-mail toninelli@lens.unifi.it)

**Author Affiliations:**

[1]European Laboratory for Nonlinear Spectroscopy (LENS), 50019 Sesto Fiorentino, Italy

[2]National Institute of Optics (CNR-INO), 50125 Florence, Italy

[3]Dipartimento di Fisica ed Astronomia, Università degli Studi di Firenze, 50019 Sesto Fiorentino, Italy

[4]Centre for Quantum Science and Technology in Arcetri (QSTAR), 50125 Florence, Italy

[5]Institute of Nanotechnology, Karlsruhe Institute of Technology (KIT), Eggenstein-Leopoldshafen 76344, Germany

[6]Institute of Physics, University of Muenster, 48149 Muenster, Germany

[7]The Blackett Laboratory, Department of Physics, Imperial College London, London SW7 2AZ, United Kingdom

[8]Laboratory of Nano-Optics, University of Siegen, 57072 Siegen, Germany

# Beaming light from a quantum emitter with a planar optical antenna

Simona Checcucci,[1,2,3,4] Pietro Lombardi,[1,2,3] Sahrish Rizvi,[1] Fabrizio Sgrignuoli,[1,3] Nico Gruhler,[5,6] Frederik B. C. Dieleman,[7] Francesco Saverio Cataliotti,[1,3,4] Wolfram H. P. Pernice,[6] Mario Agio,[1,2,4,8] and Costanza Toninelli[1,2,4]

[1] European Laboratory for Nonlinear Spectroscopy (LENS), 50019 Sesto Fiorentino, Italy

[2] National Institute of Optics (CNR-INO), 50125 Florence, Italy

[3] Dipartimento di Fisica ed Astronomia, Università degli Studi di Firenze, 50019 Sesto Fiorentino, Italy

[4] Center for Quantum Science and Technology in Arcetri (QSTAR), 50125 Florence, Italy

[5] Institute of Nanotechnology, Karlsruhe Institute of Technology (KIT), Eggenstein-Leopoldshafen 76344, Germany

[6] Institute of Physics, University of Muenster, 48149 Muenster, Germany

[7] The Blackett Laboratory, Department of Physics, Imperial College London, London SW7 2AZ, United Kingdom

[8] Laboratory of Nano-Optics, University of Siegen, 57072 Siegen, Germany

**Corresponding authors:**

Correspondence and requests for materials should be addressed to Mario Agio (e-mail: mario.agio@uni-siegen.de, phone number: +49 271 740 3532) and Costanza Toninelli (e-mail: toninelli@lens.unifi.it, phone number: +39 055 457 2134, fax number: +39 055 457 2451).

# Abstract

The efficient interaction of light with quantum emitters is crucial to most applications in nano and quantum photonics, such as sensing or quantum information processing. Effective excitation and photon extraction are particularly important for the weak signals emitted by a single atom or molecule. Recent works have introduced novel collection strategies, which demonstrate that large efficiencies can be achieved by either planar dielectric antennas combined with high numerical-aperture objectives or optical nanostructures that beam emission into a narrow angular distribution. However, the first approach requires the use of elaborate collection optics, while the latter is based on accurate positioning of the quantum emitter near complex nanoscale architectures; hence, sophisticated fabrication and experimental capabilities are needed. Here, we present a theoretical and experimental demonstration of a planar optical antenna that beams light emitted by a single molecule, which results in increased collection efficiency at small angles without stringent requirements on the emitter position. The proposed device exhibits broadband performance and is spectrally scalable, and it is simple to fabricate and therefore applies to a wide range of quantum emitters. Our design finds immediate application in spectroscopy, quantum optics and sensing.

# Introduction

Experiments at the single emitter level require a highly efficient interface with photons. Integrated optics offer various opportunities based on the strong field confinement and resonant effects in nanostructures[1,2,3]. Optical nanoantennas in planar configurations, based on dielectric or metallo-dielectric multilayers structures[4,5,6] and high numerical-aperture (NA) setups, have been recently investigated with the purpose of optimizing the collection efficiency from quantum emitters. However, alternative approaches are intensely being pursued to achieve free-space collection and flexibility in addressing single emitters. As a matter of fact, these conditions are paradigmatic for the development of single-photon sources in quantum optics[7,8], fluorescence-based microscopy[9] and spectroscopy[10], as well as for sensing applications[11]. Scanning optical microcavities, for example, provide enhanced light-matter interaction but suffer from a limited spectral bandwidth[12,13,14]. In principle, tightly focused light can lead to large coupling strengths[15,16], but sophisticated optical setups are required because a single emitter, which radiates as a Hertzian dipole, has an inherently weak directionality. However, optical antennas can be engineered to modify the radiation pattern over a broad spectral range[17]. Therefore, they can be exploited to reduce the NA of the collection/excitation optics without compromising the efficiency and the bandwidth. In this context, several nanoarchitectures have been proposed recently, including optical Yagi-Uda antennas[18,19,20], traveling wave antennas[21,22], and patch antennas[23,24,25]. Nanometric control of both the geometry during the fabrication process and the relative position of the quantum emitter is required. Furthermore, the resonant nature or the three-dimensional complex geometry means that optical antennas are not readily applicable for every light source.

Here, we demonstrate a much simpler antenna configuration that overcomes all these challenges and simultaneously improves the directionality of the emission at small polar angles and the operational bandwidth. Specifically, we experimentally show funneling of single molecule radiation at 800 nm into a narrow lobe with a semi-angle of 20° and a spectral bandwidth of 80 nm. These features promote our design for the coupling of single emitters to external optical channels with low NA.

## Materials and Methods

We investigate the emission properties of a quantum emitter in different environments; we begin with a Hertzian dipole in a homogeneous medium. Then, a glass plate is introduced underneath to define the experimental reference configuration. In our first design, a single molecule is placed near a highly reflecting mirror (see Figure 1a), and then, it is embedded in a double-element planar antenna (see Figure 1b). A strong beaming effect emerges in the radiation pattern, which is analyzed theoretically and then experimentally by back focal plane (BFP) measurements of single-molecule fluorescence in the corresponding multilayer structures. Complementary data, such as fluorescence lifetime, saturation intensity and collected single-photon count rates, are provided in order to clarify the underlying physical mechanism. Further calculations, obtained as a function of the emission wavelength and geometrical parameters, highlight the robustness and broadband nature of the proposed device.

In this section, we briefly introduce the model for an oscillating dipole inside a planar multilayer structure, then discuss the details of the fabrication procedures and finally describe the experimental setup to characterize our samples and the performance of our planar directional antenna.

The radiation pattern of a Hertzian dipole embedded in an arbitrary multilayer structure was computed in order to design our antenna structures, as shown in Figures 1a and 1b. The electromagnetic field is decomposed into plane and evanescent waves. Generalized Fresnel coefficients for transmission and reflection enforce the boundary conditions between each planar interface of the multilayer. The source field is thus propagated outside the multilayer structure and it can be used to compute the radiation pattern[26,27], as shown in Figures 1c and 1d. For comparison with the experimental results, where emission from a single molecule in the objective back focal plane (BFP) is recorded, the transmitted power calculated onto a hemisphere in the far field is multiplied by $1/\cos\theta$ (the inverse squared of the apodization function of an aplanatic system). Furthermore, we obtained the total transmitted ($P_{rad}$) and emitted ($P_{tot}$) powers by integrating the corresponding power densities as a function of the in-plane wavevector[26,27]. This allows us to identify dissipation channels and to quantify the antenna efficiency, which is defined as $\eta = P_{rad}/P_{tot}$.

Our samples were obtained by embedding a 50-nm-thick anthracene (Ac) layer containing isolated dibenzoterrylene (DBT) molecules within dielectric spacers and metal layers, as exemplified in Figure 2a. This host-matrix combination (DBT-Ac) has been extensively studied, and it exhibits long photostability at room temperature, emission in the near infrared (zero-phonon line at 785 nm), high brightness, and controlled dipole orientation and film thickness[28]. All the samples were fabricated through the subsequent superposition of thin layers. The DBT-Ac layer and the spacer layers, namely, hydrogen silsesquioxane (HSQ) and polyvinyl alcohol (PVA) were deposited by spin-coating, while the gold layers were deposited by sputtering and physical vapor deposition. The thickness of each layer was investigated by interferometric microscopy (PVA) or atomic force microscopy (HSQ, Ac, gold). The DBT-Ac layer was prepared using the same procedure described in Ref. 28. PVA layers were prepared starting with water solutions of various concentrations that were then spin-coated at different velocities in order to obtain the desired thicknesses. For instance, 3% PVA was spin-coated at 5000 rpm for 120 s, with a ramp of 5 s at 1500 rpm, to obtain a thickness of 70 ± 4 nm. The thin films deposited were dried at ambient conditions. A HSQ resist with 2% concentration was used and spin-coated at 1500 rpm for 60 s, and then a baking procedure with a hotplate (150° C for 2 min and then 200° C for 8 min) and oven (400° C for one hour) allows us to reach a thickness of approximately 60 nm. By dry etching in Ar/O2 plasma, the film was thinned to a thickness of 50 ± 3 nm, which was confirmed by optical reflectometry.

The emission profile was investigated at room temperature with the setup illustrated in Figure 2b. A continuous wave (CW) diode laser at 767 nm and a pulsed Ti:sapphire laser (50 fs-long pulses) were used for the excitation of the DBT molecules. A dichroic mirror (DM) and a long-pass (LP) filter in the collection path allow for the detection of the weak red-shifted fluorescence signal. The wave-vector distribution of the emitted light has been studied via BFP imaging[29]. In particular, the actual system parameters result in an angular resolution that is equal to 4° in the worst case, i.e., at the edge of the NA. A conversion from the electron-multiplier charge-coupled device (EM-CCD) pixels into degrees was performed in order to make a comparison between the theoretical analysis and the experimental results. In the latter, we took into account the response function of the optical system. The setup is also equipped with a spectrometer, a Hanbury Brown-Twiss (HBT) module and fast avalanche photodiodes (APDs) for the measurement of the emission spectrum, the single-photon statistics, and the lifetime and saturation curves, respectively.

## Results and discussion

The radiation pattern of a Hertzian dipole in a homogeneous infinite medium is illustrated in Figures 1c and 1d by the black dotted curves. The dipole orientation is parallel to the object plane of a microscope objective and the power has been integrated over the azimuthal angle. It is well known that 50% of the emission propagates in the opposite direction away from the observer and a significant amount of power is radiated below an angle of 90°, which even lies outside the collection angle of a high-NA objective. Therefore, the first step in our design has been to consider the modification of the radiation pattern when a metallic mirror, denoted as the reflector, is placed below the Hertzian dipole, which is represented by a black arrow in Figure 1a. We identify three possible configurations as a function of the distance $d_1$ between the emitter and reflector. For $d_1 \ll \lambda/n$, where $\lambda$ is the emission wavelength and $n$ the refractive index of the medium, the power is mostly transferred to the metal by absorption or excitation of surface plasmon polaritons, or the emission is suppressed by destructive interference with the image dipole, by assuming the metal is a perfect mirror[30]. In both cases, quenching of the emission is observed. In the opposite extreme case, for $d_1 \gg \lambda/n$, multiple interference, which occurs between the light directly emitted by the Hertzian dipole and the light reflected by the metal, gives rise to an emission pattern that may present various lobes and is hence unsuitable for efficient light collection. In our work, we focus on the intermediate values of $d_1$, for which we have calculated the radiation pattern and the total emitted power with a semi-analytical approach, as described in the Materials and Methods section.

The introduction of a reflector element substantially directs the emission of a Hertzian dipole, as shown in Figure 1c for $d_1$ = 80 nm ~ $\lambda/(6n)$, $\lambda$ = 785 nm and $n$ = 1.5. Intuitively, these results can be explained in terms of the image dipole induced in the reflector by the source. When the two dipoles radiate with an appropriate phase difference, which depends on $\lambda/n$, $d_1$ and the optical constants of the reflector, they constructively interfere in the forward direction, which results in a beaming effect[31]. For a strongly reflective mirror, such as a 100 nm-thick gold film in the near infrared[32], we have found that beaming occurs in the range of $\lambda/6n \leq d_1 \leq \lambda/4n$. For $\lambda$ = 785 nm and $n$ = 1.5, this range spans approximately 50 nm, which can be easily controlled with state-of-the-art technology.

The directionality can be improved by placing a second semi-reflective film at an appropriate distance $d_2$ from the Hertzian dipole (see Figure 1b and 1d), which functions as a passive antenna element. In practice, the field emitted by the dipole induced in such an element constructively interferes with that of the source in the forward direction and beams the radiation pattern in a similar manner as a director does in a Yagi-Uda configuration[31]. We have found that with a 20-nm thick gold film, the emission is channeled into a single lobe, which has a width with a semi-angle of approximately 24°, as shown in Figure 1d, when $d_1$ = 130 nm and $d_2$ = 70 nm. We remark that this occurs even for a randomly oriented dipole, mainly because the two dipole orientations in the plane are equivalent for a planar structure. Investigating the radiation pattern as a function of the thickness of the director, we find that for very thin films, i.e., for films thinner than approximately 5 nm, or for films thicker than 40 nm, the antenna performance worsens either because the beaming effect disappears or the transmission through the gold film becomes extremely small (see the Supplementary Information).

To quantify the efficiency of our structures, in Figure 1e, we plot the collected power integrated over the azimuthal coordinate ($P_{coll}$) as a function of the collection angle. There is a substantial gain in $P_{coll}$ with respect to the homogeneous case (black dotted curve), especially for the configuration with the reflector and detector (blue solid curve) and for small angles. Moreover, this holds for a randomly oriented dipole (blue dashed curve). All curves in Figure 1 are normalized to the total radiated power in a homogeneous infinite medium with a refractive index of 1.5 ($P_{hom}$). When the collection angle is 90°, $P_{coll}/P_{hom}$ reaches a value of 0.6 for the case with only a reflector (see Figure 1a) and approximately 1.2 for the case with the reflector and director (see Figure 1b). Deviations of $P_{coll}/P_{hom}$ at 90° from unity primarily occur because the Hertzian dipole experiences a modification of the radiative rate, which will be analyzed. We remark that our designs represent an improvement in collection efficiency by more than one order of magnitude when considering the coupling to standard multimode fibers, for which NA = 0.4 is typical. Likewise, the excitation efficiency is improved and, by matching the spatial profile of the emitter with that of a focused laser beam, the antenna allows for lower pump powers, which in turn implies that a better signal to noise ratio is achieved.

At microwave frequencies, a metallic grid placed near a ground plate turns a Fabry-Perot cavity into an antenna with high directionality[33]. Scaling such structures to optical wavelengths would not be trivial due to the excitation of surface plasmon-polaritons in hole arrays, which affects the light emission and its spatial redistribution[34,35]. On the contrary, the element director in our geometry is a homogeneous thin film, which makes the lateral position of the dipole irrelevant, prevents complex plasmonic effects and precludes the need for nanofabrication tools. This is also in contrast to the typical optical Yagi-Uda antenna, where the position of the quantum emitter must be controlled with nanometer accuracy with respect to the feeding element[18,36]. Furthermore, we note that our antenna performance is only weakly dependent on the wavelength, and hence, it cannot correctly be described using a cavity effect[37]. Finally, the constructive interference, which gives rise to the beaming effect, corresponds to a non-resonant condition for the equivalent cavity (we operate more than a FWHM away from resonance, which is calculated from the transmission of a plane wave under normal incidence). Conversely, no beaming of the emission pattern is predicted at resonance, as determined by numerical simulations reported in the Supplementary Information.

The reference for our experiments is a 50-nm-thick DBT-Ac layer deposited on a glass cover slip. The emission spectrum, measured with a standard spectrometer, is shown in Figure 2c. The zero-phonon-line (ZPL) is found at λ = 785 nm and the spectrum exhibits a full-width at half-maximum (FWHM) of approximately 50 nm. The second-order intensity autocorrelation function, which is measured with a Hanbury Brown-Twiss setup, provides evidence of single-molecule emission, with a signal suppression of 98% at zero delay (as observed in Figure 2d, where no background subtraction has been applied). With a standard time-correlated single photon counting scheme, time-resolved fluorescence decay curves are obtained, such as the curve plotted in Figure 2e, and hence, from a single exponential fit, the excited-state lifetime of the molecule is estimated. A statistical analysis of the lifetime distribution in the reference can be found in previous work[38], which yields an average value of 4.1 ± 0.4 ns for the excited-state lifetime. In the DBT-Ac system, the molecule is oriented parallel to the Ac film, which simplifies the comparison between the BFP measurements and the theoretical predictions.

The main results are presented in Figure 3, where the upper panels correspond to experimental BFP images integrated over the emission spectrum and the bottom panels correspond to calculations obtained for λ = 785 nm. The corresponding structures, which are sketched in the top part of Figure 3, from left to right are as follows: (Reference) a 50 nm-thick DBT-Ac film deposited on top of a glass cover slip; (Single mirror) a 50 nm-thick DBT-Ac film separated from a 100 nm-thick gold film by a 70 nm-thick PVA layer; (Single mirror 2) the previous configuration with a 300 nm-thick PVA layer; (Double mirror) the antenna depicted in Figure 2b with a glass cover slip followed by a 20 nm-thick gold film, an annealed 50-nm-thick HSQ layer, a 50 nm-thick DBT-Ac layer, a 70 nm-thick PVA layer and 100 nm-thick gold film. Note that all layers between the reflector and director in this configuration have very similar refractive indices. All thicknesses refer to growth parameters that were verified using atomic force microscopy and interferometric techniques (see the Materials and Methods section for details). Fluctuations can occur, but they have not significantly influenced the emission, which proves the robustness of the beaming effect. Therefore, the parameters relevant to the double mirror configuration have been selected in such a way that beaming is observed for most of the investigated molecules, although it is dispersed at different positions inside the host matrix. However, this geometry does not maximize the total emitted power, which is discussed later with respect to Figure 4c.

The experimental results are compared with our model along linear cuts taken in the direction perpendicular to the dipole orientation (see Figures 3e-h). The theoretical data are shown for two extreme cases with the DBT molecule lying 2 nm below the top or above the bottom surface of the Ac layer, which are represented by the green and blue curves, respectively. The double-lobe characteristic of a horizontally oriented dipole placed in close proximity to a dielectric interface[39] is visible in Figures 3a, 3e and 3i. The addition of a reflector can create completely different radiation profiles. For a PVA thickness of 70 nm, Figures 3b, 3f and 3j show that the emission pattern is confined in one central lobe. However, Figures 3c, 3g and 3k show that for a PVA thickness of 300 nm, most of the emission falls outside the NA of the objective (1.4), which is represented by a white circle. Finally, the antenna with the reflector and director is presented in Figures 3d, 3h and 3l. In this case, the emission from a single DBT molecule in our planar optical antenna amounts to a semi-angle of only 20° at half maximum. This corresponds to a collection efficiency of 43% at 24° and 90% at 56°.

In all four cases, we observe good agreement between the experimental findings and the theoretical calculations, while assuming nominal parameters for the layer thicknesses and considering the uncertainty in the position of the DBT molecule within the Ac crystal, as well as the transfer function of the setup. In particular, a step truncation in any of the Fourier planes, which is given by the finite size of the objective, corresponds to smoothing and the occurrence of ripples at the edges of the NA. The anti-reflection coating of the objective exhibits strong wavelength dependence at large angles, as shown in the Supplementary Information. The reported curves are not corrected for the transfer function of the setup, and the invalid region is shaded in grey. Overall, we can reliably compare the theoretical and experimental results up to approximately 55°, which is the angle at which the transmission reduces by a factor of two. Note that the apparent discrepancy between theory and experiment for the Single mirror 2 configuration is due to the normalization values, which differ because of the maximum measurable angle. All the experimental measurements fall within the theoretical curves that account for the possible positions of the DBT molecule in the Ac film.

The robustness and versatility of the directive antenna analyzed in Figure 3d has been verified by performing additional theoretical and experimental analysis. First, we characterized the spontaneous emission rate of the DBT molecule to rule out coupling to cavity-like modes of the structure with the reflector and director, as well as to quantify enhancement or quenching of the fluorescence. Statistics of the lifetime measurements have been collected and compared to an experimental reference sample[38]. The average value and standard deviation from the data set reported in Figure 4a yield a lifetime of 3.8 ± 0.4 ns for DBT in the antenna configuration, and this value is perfectly comparable to that of the reference. From a theoretical point of view, this is consistent with the calculations of the total emitted power by a Hertzian dipole ($P_{tot}$), which is inversely proportional to the excited state lifetime and yields similar results in the two cases.

In addition, this view is confirmed by the measurement of the saturation curve (Figure 4b), which exhibits comparable values for the maximum collected power. Specifically, the count rates at saturation are 3.1 ± 0.5 Mcps for the reference sample of Figure 3a and 2.8 ± 0.7 Mcps for the directional antenna of Figure 3d. This is consistent with comparable decay rates and similar losses in both cases. For the reference sample, the collection efficiency is limited by the objective NA; however, in the antenna structure, we have calculated an efficiency on the order of 55%, which varies as a function of the dipole position and multilayer geometrical parameters.

The excitation efficiency is sensitive to the mode matching conditions. As a consequence of reciprocity, the radiation pattern of an antenna is equal to its receiving pattern. The relationship holds here because of the broadband performance of our design. Moreover, the blue-shifted excitation experiences a small field enhancement in the antenna. Hence, the interaction with the molecule has improved due to better mode matching and field concentration. In particular, we report a saturation power ($P_s$) that is a factor of 12 smaller for molecules within the antenna of Figure 3d compared to the reference sample of Figure 3a, i.e., $P_s$ decreases from 32 ± 8 µW to 2.6 ± 0.7 µW. Other emitter configurations would result in different combinations of beaming and excitation enhancement. A certain flexibility in finding an optimal trade-off stems from the fact that many configurations yield similar radiation patterns (see the Supplementary Information). All our data confirm the view that the modified emission profile yields an increased coupling efficiency, e.g., with a weakly focused laser beam or an optical fiber.

In Figure 4c, we analyze the antenna efficiency ($P_{rad}/P_{tot}$) by considering $P_{tot}$ and $P_{rad}$ as a function of the dipole distance to the reflector (blue and red solid curves, respectively). Here, the overall sample thickness is kept constant ($d_1 + d_2$ = 200 nm), and the geometrical parameters are those of Figure 1b. In particular, the antenna working region is defined around $d_1$ = 130 nm, and it corresponds to an intensity that is approximately 70% higher than the intensity for a dipole in a homogeneous medium with the same refractive index $n$ = 1.5. The antenna efficiency for this set of parameters is 65%. $P_{rad}/P_{tot}$ reaches 70% at approximately $d_1$ = 80 nm and it remains above 50% for $d_1$ values between 30 nm and 160 nm, which largely shows that the efficiency of our antenna is large and weakly dependent on the dipole position. For applications where losses are critical, different materials and designs must be devised. In general, absorption losses, the presence of guided modes in dielectric layers or the excitation of surface plasmon polaritons reduce the antenna efficiency, as shown by the steep increase of $P_{tot}$ when the dipole approaches the metal layers. To stress this point, in the same figure, we plot the power absorbed by the reflector ($P_{abs}$, blue dashed curve), computed in the quasi-static approximation, for the configuration of Figure 1a (Single mirror). These effects can also be observed in the power densities displayed in the Supplementary Information. Lastly, we notice that the power radiated by a Hertzian dipole near a perfect mirror ($P_{rad}$, red dashed curve) is compatible with the calculations for $P_{coll}$ in Figure 1e, which results in approximately 60% of $P_{hom}$ for collection up to 90°. This illustrates another mechanism for the reduction of $P_{coll}$: when the source is near the reflector (or the director), the induced image dipole tends to decrease $P_{tot}$, which is equal to $P_{rad}$ for an ideal metal. Therefore, we remark that reported single-molecule lifetime experiments near a silver mirror have been in excellent agreement with the classical theory[40]. We deduce that the high intensity collected at small angles from our optical antenna stems from the directional emission rather than from an overall fluorescence enhancement.

Finally, in order to stress the robustness of the proposed device, as a figure of merit, we considered $P_{coll}/P_{rad}$ within an angle of 24°, which corresponds to the NA of a standard multimode optical fiber. We found a wide range of dipole positions in which the efficiency is larger than 60%, even when $P_{rad}$ slightly decreases (e.g., as observed in Figure 4c for values of $d_1$ smaller than 100 nm). We also investigated $P_{coll}$ as a function of $d_1$ and $d_2$ for a 20 nm-thick gold director. We observe that more than one configuration leads to similar results, and in particular, whenever the total antenna length is increased by integer multiples of $\lambda/2n$. For example, for $d_1$

~ 130 nm and $d_2$ ~ 330 nm the directionality is even higher, and the radiation pattern has a semi-angle of 12° at half maximum. In contrast, $P_{coll}/P_{hom}$ is only 0.5 in this case. Given $n$ = 1.5 and collection in air, a total length larger than approximately 1.5 µm leads to a smearing out of the beaming effect because of the presence of lobes at higher angles (see the Supplementary Information). The large bandwidth of the directive antenna can be observed from Figure 4d, where $P_{coll}/P_{hom}$ is reported as a function of the wavelength. The antenna bandwidth, which we define as the FWHM of $P_{coll}/P_{hom}$ for $\theta$ = 24°, can be estimated from the red curve in Figure 4d, and it is approximately 80 nm at approximately 785 nm.

We remark that the beaming effect can be understood as a first approximation of the interference between the Hertzian dipole and its image charges, which are induced in the reflector and director elements. In particular, we have targeted the constructive interference under small angles in the outgoing direction from the director. This condition does not maximize the field between the two mirrors, as in the case of a resonant cavity. We first established a minimum distance between the dipole and the metal layers to exclude quenching effects and performed calculations starting with an educated guess of a distance of approximately $\lambda/4n$ from the reflector, which maximizes forward scattering. This sets a range of values for the dipole position from the reflector. Then, we scanned the overall reflector-director distance and recorded both the emission pattern as well as the radiation efficiency to establish the optimal conditions. There is clearly a trade-off between a weak modulation of the density of states, which affects $P_{tot}$, the efficiency ($P_{rad}/P_{tot}$), as inferred from Figure 4c, and the modified radiation pattern, which determines the fraction of $P_{rad}$ collected within a given solid angle. Although our geometry is very similar to a Fabry-Pérot cavity, we would like to note the following observations. For the configuration in Figure 4d, we observed a threefold maximum enhancement of $P_{tot}$ with respect to an infinite medium and a broad resonance with a FWHM larger than 100 nm (blue dashed curve). Interestingly, by calculating the power density as a function of the wave vector (see Supplementary Information), one observes that it contains the contribution of surface plasmon polaritons, which corresponds to approximately half of $P_{tot}$. However, the Purcell enhancement of the cavity would not correctly describe the operation, as we work more than a FWHM away from the resonant condition. Our design, which maximizes the constructive interference in the output direction around small angles, is clearly distinct from the design of resonant antennas, which maximizes the field inside the structure or analogously $P_{tot}$ and $P_{rad}$. Such a difference can be perceived, for instance, by comparing the spectral position of the maxima in Figure 4d for the total collected power (blue curve) and that collected within 24° of the semi-angle θ (red curve). Further insight is provided in the Supplementary Information, where the near-field intensity is displayed for the structure that exhibits wave beaming effects and shown at the resonant wavelength of the equivalent cavity. Clearly, no beaming effect is observed for a resonant dipole. Finally, both theoretical and experimental investigations show neither appreciable enhancement nor reduction in the emission rate compared to that of the reference. We thus conclude that when beaming occurs, the density of states is only weakly altered by the presence of the mirror and director elements.

## Conclusions

We have proposed a planar optical antenna design that strongly directs the radiation of quantum emitters. We have discussed the conditions that give rise to such an effect and provided experimental evidence of our theoretical findings, which shows that the beaming of light from a single molecule in a semi-cone is only 20° wide. The emission pattern resembles a weakly focused Gaussian laser beam or the radiation profile of a multimode fiber. This can significantly improve the collection and excitation of quantum emitters. The operating principle is similar to that of a Yagi-Uda antenna[31], where in place of resonant elements, such as nanoparticles[18,36], we make use of the dipoles induced in the reflector and director, which are made of thin metal films. This leads to a significant simplification of the structure and a much higher tolerance in selecting the geometrical parameters. In addition, the measured radiation patterns are narrower than in recent experimental demonstrations based on more sophisticated optical antennas[18,21,23,41]. Furthermore, our device is compatible with collection in free space, and it can be easily applied to other quantum emitters, such as fluorescent dyes[42], quantum dots[43], and color centers[44], or scaled to other wavelengths. Because of its performance and simplicity, it is likely that such a design will immediately find broad applications in spectroscopy, sensing and quantum optics.

## Acknowledgments

The authors would like to thank V. Greco and A. Sordini for measuring the thickness of the thin films, B. Tiribilli, F. Dinelli and A. Flatae for inspection of the samples by atomic force microscopy, D.S. Wiersma for access to clean room facilities, M. Bellini and C. Corsi for Ti:sapphire operation, G. Mazzamuto for technical help and discussion and M. Gurioli for stimulating discussions. This work benefited from the COST Action MP1403 "Nanoscale Quantum Optics," which is supported by COST (European Cooperation in Science and Technology).

## Competing Interests

M.A., S.C., P.L., S.R., F.S. and C.T. have filed the patent application "Device for the beaming of light emitted by light sources, in particular fluorescence of molecules," Patent pending PCT/EP2016/058069 filed on 13.04.16.

## Author Contributions

M.A. and C.T. conceived, planned and supervised the project. F.S. performed preliminary numerical simulations of the structure with the reflector. S.C. theoretically investigated and designed the structures with the semi-analytical approach described in the Methods section. S.C. and S.R. fabricated the samples. N.G. and W.P. provided glass cover slips with thin gold films and HSQ. S.C., P.L. and S.R. performed back focal plane, autocorrelation and saturation measurements, with the help of F.D. for the lifetime experiments.

M.A., S.C., P.L. and C.T. analyzed the data and discussed them with all authors. M.A., S.C. and C.T. wrote the manuscript with feedback from all authors.

Supplementary Information accompanies the manuscript on the "Light: Science & Applications website (http://www.nature.com/lsa/)."

# List of Figures:

**Figure 1. Concept of a planar directional antenna and theoretical analysis.**

(**a-b**) Sketch of a planar directional antenna: the induced image-dipoles are identified by the red and the blue arrows in the reflector and in the director, respectively, while the source dipole is represented by a black arrow. The emitter is placed at distances $d_1$ and $d_2$ from the reflector and the director, respectively. In our study, the reflector and the director are 100-nm-thick and 20-nm-thick gold films, respectively. The medium above the reflector has a refractive index $n$ = 1.5, while a semi-infinite top layer of air is considered ($n$ = 1). (**c-d**) Radiation pattern integrated over the azimuthal coordinate as a function of the polar angle θ of a horizontal Hertzian dipole (HD), emitting at λ = 785 nm, coupled to the planar directional antenna of Figure 1a ($d_1$ = 80 nm) and b ($d_1$ = 130 nm and $d_2$ = 70 nm), respectively. The two curves are normalized by the maximum value of the same quantity for a dipole in a homogeneous medium with refractive index $n$ = 1.5, which is also shown by the black dotted curves. (**e**) Normalized power ($P_{coll}/P_{hom}$) as a function of the collection angle for the configurations of Figures 1a-d. The normalized power for a Hertzian dipole with random orientation (RD) is shown by the blue dashed curve. $P_{hom}$ is the total power emitted by a Hertzian dipole in a homogeneous medium of refractive index $n$ = 1.5.

**Figure 2. Experimental setup and sample characterization**.

(**a**) Sketch of the experimental setup for single-molecule spectroscopy, and BFP imaging. (**b**) Layout of the sample containing a DBT-Ac film inside the planar optical antenna. (**c**) Fluorescence spectrum of a single DBT molecule in an Ac crystalline film deposited on a glass cover slip. (**d**) For the same configuration, second-order intensity autocorrelation function measured with a Hanbury Brown-Twiss setup and (**e**) time-resolved fluorescence decay curve. From the single exponential fit (red line), one obtains the excited-state lifetime for the molecule.

**Figure 3. Experimental results and comparison with theory.**

**(a)-(d)** Sketches of the samples and corresponding normalized BFP images of a single DBT molecule acquired with an EM-CCD camera. **(e)-(h)** Comparison between calculated emission patterns for λ = 785 nm and cross-sections of the experimental BFP images. The shaded areas represent the objective transfer function, from a transmission of 0 (gray) to 1 (white). The two theoretical curves for each case correspond to different positions of the DBT molecule in a 50 nm-thick Ac film: 2 nm from the top interface (blue curve) and 2 nm from the bottom interface (green curve). Both the theoretical and experimental curves have been normalized to 1, except for the experimental curve of **(g)**, which is normalized in order to match the theoretical value at 0° (the dipole position at 2 nm). **(i)-(l)** Theoretical BFP images that reproduce the experimental results. The plots are normalized with respect to the maximum value of the radiation pattern of a Hertzian dipole in a homogeneous medium with $n$ = 1.5. The white circles show the nominal NA of the collection optics (θ = 67°). **(a), (e), (i)** Experimental reference case with glass cover slip ($n_{glass}$ = 1.52), 50 nm DBT-Ac ($n_{Ac}$ = 1.6) and air ($n_{air}$ = 1); **(b), (f), (j)** System with reflector, composed of a semi-infinite glass cover slip, 50 nm DBT-Ac, 70 nm PVA ($n_{PVA}$ = 1.49), 100 nm gold and air; **(c), (g), (k)** Same configuration as before, but with 300 nm-thick PVA layer; **(d), (h), (l)** System with director and reflector, composed of a semi-infinite glass cover slip, 20 nm gold, 50 nm HSQ ($n_{HSQ}$ = 1.4), 50 nm DBT-Ac, 70 nm PVA, 100 nm gold and air.

**Figure 4. Further analyses of antenna robustness**

**(a)** Statistical distribution of the excited-state lifetime of a DBT molecule in an Ac crystalline film, which is coupled to our planar optical antenna with the director and reflector, as shown in Figure 2b. **(b)** Typical saturation curves obtained for dipoles in the reference sample and saturation curves obtained for the dipoles, which are coupled to the double mirror antenna. For the sake of clarity, the offset and linear dependence of leakage of the pump light are subtracted. The solid lines are fits for the data, which give maximal counts for $R_{inf}$ = (2.97 ± 0.08) x $10^6$ and (2.89 ± 0.02) x $10^6$, and the saturation power $P_s$ = (49 ± 4) x 10 µW and 47.0 ± 1.4 µW, for the reference and antenna configurations, respectively. **(c)** Normalized powers associated with a Hertzian dipole at λ = 785 nm in a medium with refractive index $n$ = 1.5 at a distance $d_1$ from a reflector. The distance between the 100 nm-thick gold reflector and a 20 nm-thick gold director is kept constant at 200 nm. The blue solid curve refers to the total power emitted by the dipole in the antenna ($P_{tot}/P_{hom}$), whereas the red solid curve represents the power transmitted (radiated) in the upper medium above the director ($P_{rad}/P_{hom}$). The blue dashed curve represents power absorption ($P_{abs}/P_{hom}$) by energy transfer to the reflector, while the red dashed curve represents $P_{tot}/P_{hom}$ for the case where the Hertzian dipole is in front of a perfect mirror. The media below the reflector and above the director are air and glass, respectively. **(d)** Normalized collected power $P_{coll}/P_{hom}$ up to θ = 24° (red solid curve) and θ = 90° (blue solid curve) as a function of wavelength for $d_1$ = 130 nm and the antenna parameters of Figure 4c, except that the external media are interchanged. $P_{coll}/P_{hom}$ up to θ = 90° is also shown for a Hertzian dipole in glass (black dotted curve) or 130 nm from a gold reflector (red dashed curve).

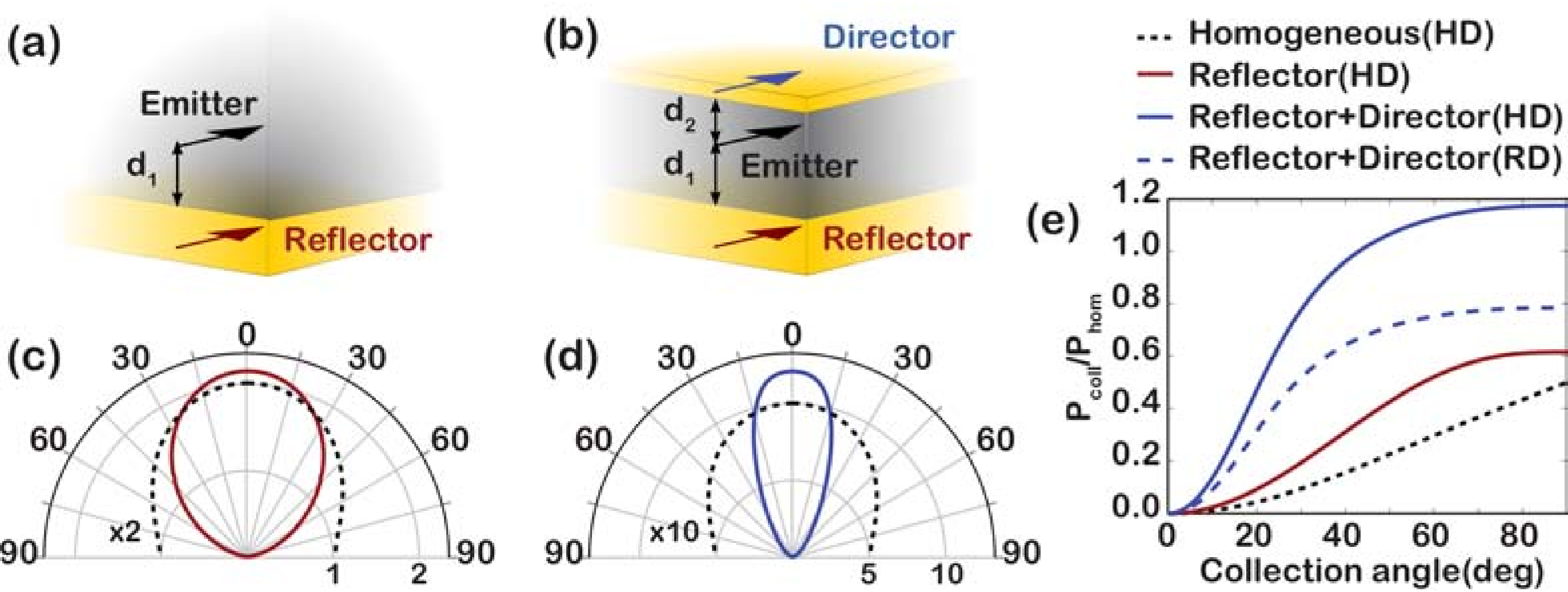

(a) HSQ PVA DBT-Ac Sample

Reference measurements

(b)

(c) Intensity (a.u.) vs Wavelength (nm), 760–840

(d) Norm. coinc. vs Time (ns)

(e) Norm. counts vs Time (ns)

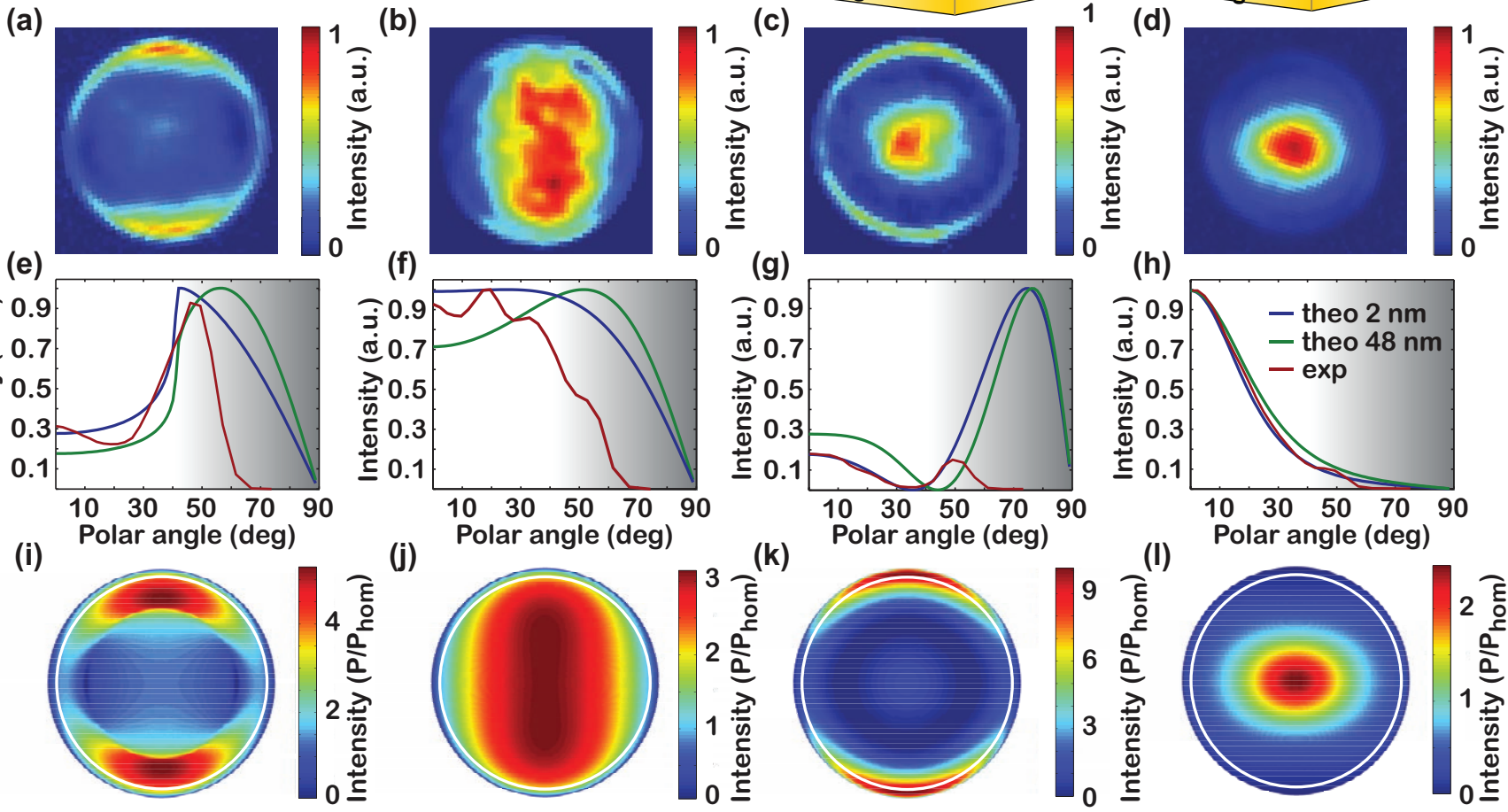

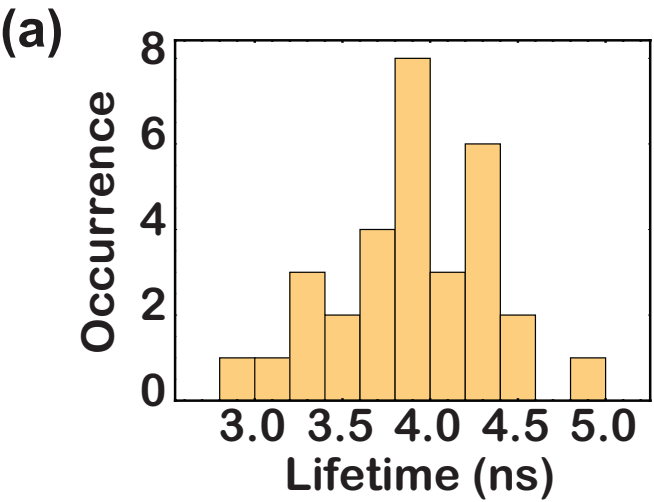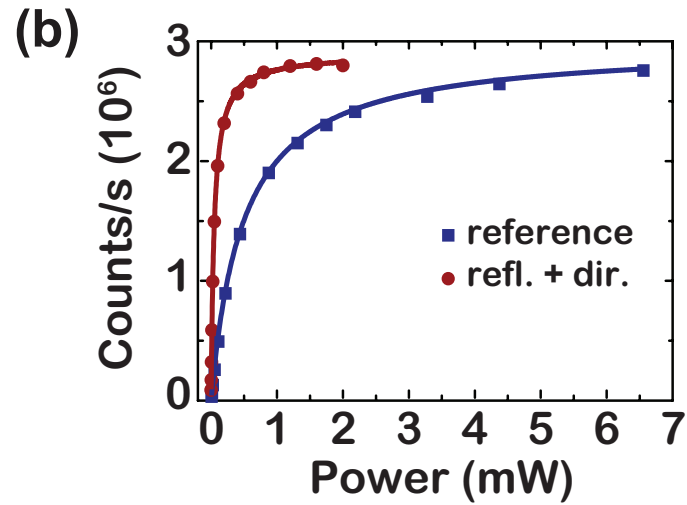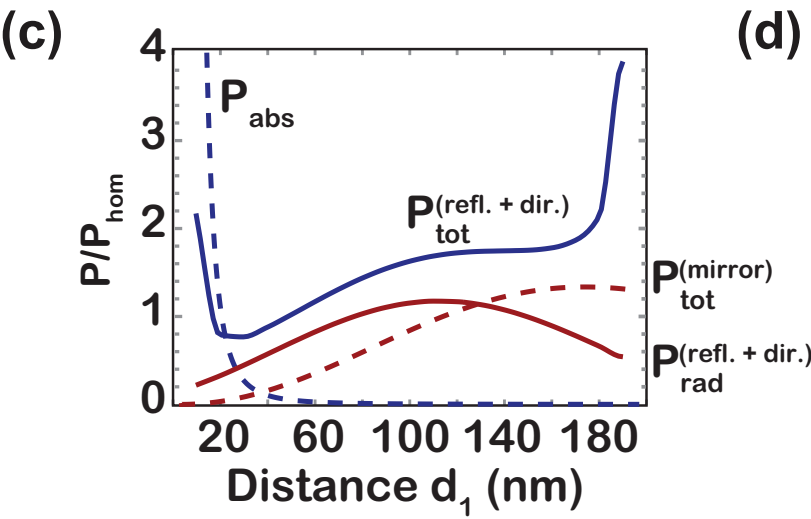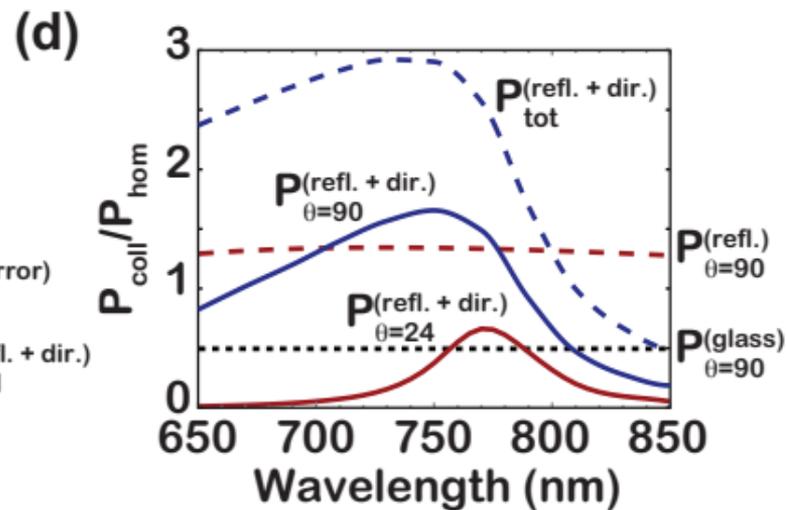